\documentclass[12pt]{article}
\pdfoutput=1

\usepackage{amsmath}
\usepackage{amsfonts}
\usepackage{amscd}
\usepackage{amsthm}
\usepackage{setspace}

\usepackage{graphicx}
\usepackage{authblk}
\usepackage{caption}
\usepackage{ytableau}
\usepackage{mathtools}

\usepackage[OT2,T1]{fontenc}
\DeclareSymbolFont{cyrletters}{OT2}{wncyr}{m}{n}
\DeclareMathSymbol{\Sha}{\mathalpha}{cyrletters}{"58}

\def\Z{\mathbb{Z}}

\def\P{\mathbb{P}}

\def\til{\tilde}

\begin{document}

\begin{titlepage}

\begin{flushright}
KEK-TH 2030 
\end{flushright}

\vskip 1cm

\begin{center}

{\large K3 surfaces without section as double covers of Halphen surfaces, and F-theory compactifications}

\vskip 1.2cm

Yusuke Kimura$^1$
\vskip 0.4cm
{\it $^1$KEK Theory Center, Institute of Particle and Nuclear Studies, KEK, \\ 1-1 Oho, Tsukuba, Ibaraki 305-0801, Japan}
\vskip 0.4cm
E-mail: kimurayu@post.kek.jp

\vskip 1.5cm
\abstract{We construct several examples of genus-one fibered K3 surfaces without a global section with type $I_{n}$ fibers, by considering double covers of a special class of rational elliptic surfaces lacking a global section, known as Halphen surfaces of index 2. The resulting K3 surfaces have bisection geometries. F-theory compactifications on these K3 genus-one fibrations without a section times a K3 yield models that have $SU(n)$ gauge symmetries with a discrete $\Z_2$ symmetry.}  

\end{center}
\end{titlepage}

\section{Introduction}
\par There are a number of advantages of using the F-theory approach for model building in particle physics. SU(5) GUT is naturally realized in F-theory with matter in SO(10) spinor representations. The problem of weakly coupled heterotic string, which is addressed in \cite{Witten}, can also be avoided. Local F-theory model buildings \cite{DWmodel, BHV1, BHV2, DWGUT} have been emphasized in recent studies. Global aspects of F-theory compactifications, however, need to be analyzed eventually to address the issues of gravity and the early universe, including inflation. F-theory/heterotic duality \cite{Vaf, MV1, MV2, Sen, FMW} particularly provides insights into global models of F-theory compactifications. F-theory/heterotic duality, which states that F-theory compactified on a K3 fibered Calabi--Yau $(m+1)$-fold is equivalent to heterotic string theory compactified on a torus fibered Calabi--Yau $m$-fold, is formulated when the stable degeneration limit \cite{FMW, AM} of a K3 fibration is considered \footnote{See, e.g., \cite{AHK, BKW, BKL, CGKPS, MizTan, KRES} for recent progress on stable degenerations in F-theory/heterotic duality.} on the F-theory side.  
\par In recent studies, F-theory compactifications on genus-one fibrations without a section are considered. One of the reasons such compactifications are studied is that a discrete gauge symmetry \footnote{Recent progress in discrete gauge symmetries can be found, e.g., in \cite{KNPRR, ACKO, BS, HSsums, CIM, BCMRU, BCMU, HS, BRU, KKLM, BGKintfiber, HS2, GPR, CGP}.} arises in F-theory models on genus-one fibered spaces lacking a global section, as discussed in \cite{MTsection}. Genus-one fibered Calabi--Yau manifolds $M$ whose Jacobian fibrations $J(M)$ are identical form a group, which is referred to as the Tate--Shafarevich group $\Sha(J(M))$. As discussed in \cite{MTsection}, the discrete part of this group $\Sha(J(M))$ is identified with the discrete gauge symmetry that arises in F-theory compactification on the Calabi--Yau genus-one fibration $M$ without a rational section \cite{BDHKMMS}. Concretely, when Calabi--Yau genus-one fibration $M$ without a rational section admits a multisection of degree $m$, a discrete $\Z_m$ symmetry arises in F-theory compactified on the genus-one fibered Calabi--Yau manifold $M$. 
\par In this study, we construct genus-one fibered K3 surfaces without a global section, that have bisections, with type $I_n$ fiber for various $n$. Utilizing this result, genus-one fibered Calabi--Yau 3-folds may be constructed by considering fibrations of these genus-one fibered K3 surfaces over the base $\P^1$. Genus-one fibered Calabi--Yau spaces with a bisection, and F-theory compactifications on these spaces are discussed in \cite{BM, MTsection, MPTW, MPTW2, LMTW, K2, KCY4, Kdisc}. 
\par Genus-one fibrations without a global section do not admit transformation into the Weierstrass form. Owing to this, it is generally technically difficult to explicitly construct genus-one fibered Calabi--Yau manifolds lacking a rational section with specific types of singular fibers. In this note, we use rational elliptic surfaces without a global section as a tool to construct K3 genus-one fibrations lacking a global section, with a bisection, with type $I_n$ fiber for various $n$. This construction shows the existence of F-theory models with a discrete $\Z_2$ symmetry, with $SU(n)$ gauge groups. 
\par We focus on a special class of rational elliptic surfaces lacking a global section, namely, {\it Halphen surfaces of index 2}, in this note, to yield genus-one fibered K3 surfaces without a global section that have a bisection. We particularly give explicit constructions of the Halphen surfaces of index 2 with type $I_n$ fiber for various $n$. Taking double covers of these surfaces yields genus-one fibered K3 surfaces without a section, that have a bisection with type $I_n$ fiber. F-theory compactified on spaces built as the direct products of these genus-one fibered K3 surfaces without a section times a K3 surface gives four-dimensional theories with a discrete $\Z_2$ symmetry and $SU(n)$ gauge groups. Particularly, F-theory compactification on K3 genus-one fibration obtained as a double cover of Halphen surface of index 2 with type $I_5$ fiber times a K3 yields four-dimensional theory with $SU(5)$ gauge group with a discrete $\Z_2$ symmetry, which relates to GUT realization.
\par F-theory models compactified on spaces that have an elliptic fibration with a global section have been considered \cite{MorrisonPark, MPW, BGK, BMPWsection, CKP, LSW, KimuraMizoguchi}. General genus-one fibered Calabi--Yau manifolds, however, do not necessarily have a global section. Genus-one fibered Calabi--Yau manifolds that lack a global section indeed exist, and F-theory models on genus-one fibered Calabi--Yau manifolds without a section, including genus-one fibered K3 surfaces lacking a global section, have recently been studied, e.g., in \cite{BM, MTsection, AGGK, KMOPR, GGK, MPTW, MPTW2, BGKintfiber, CDKPP, LMTW, K, K2, KCY4, Kdisc}. Similar to that there are genus-one fibered K3 surfaces that lack a global section, rational elliptic surfaces without a global section are known to exist.
\par Halphen surfaces of index 2 are examples of rational elliptic surfaces lacking a global section. Discussions of the Halphen surfaces can be found in \cite{DolgachevZhang, CantatDolgachev}. Halphen surfaces of index 2 are bisection geometries. Concretely, we consider a sextic curve in $\P^2$, $f=0$, with nine simple singularities. A cubic curve, $g=0$, which passes through these nine points, always exists. We take the ratio of the sextic curve and the square of the cubic curve $[f: g^2]$; this ratio yields a projection onto $\P^1$, which results in a genus-one fibration. Exceptional divisors that arise from blowing up the nine simple singularities yield bisections. 
\par More generally, when we have a curve $\til{f}$ of degree $3n$ in $\P^2$ with nine singularities of multiplicities $n$, we obtain a Halphen surface of index $n$. Here, we considered only Halphen surfaces of index 2, $n=2$, in this study.
\par In this note, we particularly discuss Halphen surfaces of index 2 with singular fibers \footnote{The types of the singular fibers of elliptic surfaces were classified \cite{Kod1, Kod2}. Methods that determine singular fibers of elliptic surfaces can be found in \cite{Ner, Tate}.} of type $I_n$. We mainly considered situations in which sextic curve $f=0$ is reducible into several irreducible curves, such as lines, conics, and cubics. We considered the blow-up of $\P^2$ at nine points of the intersection points, among others, of these curves to obtain Halphen surfaces of index 2. The remaining intersection points, which were left un-blown up, and lines and conics (or cubics with a node) that pass through these points form an $n$-gon; this configuration yields type $I_n$ fiber. 
\par By considering double covers of Halphen surfaces of index 2 with type $I_n$ fibers, we obtained genus-one fibered K3 surfaces \footnote{These K3 surfaces belong to the K3 surfaces with involution considered in \cite{Nikfactor}.} without a section with type $I_n$ fibers. Because Halphen surfaces of index 2 have bisections, K3 surfaces obtained as double covers of Halphen surfaces of index 2 possess bisections. These K3 surfaces constructed as double covers of Halphen surfaces ramify over a fiber of Halphen surfaces. We obtained two different types of K3 surfaces depending on whether a fiber as the branching locus of the double cover is smooth or singular. 
\par We discuss F-theory compactifications on these K3 surfaces without a section, obtained as double covers of the Halphen surfaces of index 2. The correspondence of non-Abelian gauge groups that arise on the 7-branes in F-theory compactifications and the types of the singular fibers of the compactification spaces are discussed in \cite{MV2, BIKMSV}. Type $I_n$ fiber corresponds to $A_{n-1}$ singularity. In F-theory compactifications, matter representations arise from local rank 1 enhancements of the singularities of the compactification spaces \cite{BIKMSV, KV, GM, MTmatter, GM2}. Other types of matter, that arise from the structures of divisors, are discussed in \cite{KMP, Pha}. Matter fields in four-dimensional theories in the presence of a flux are analyzed using the F-theory approach in \cite{DWmodel, BHV1}.
\par Jacobian fibrations \footnote{Construction of the Jacobian of a genus-one curve is discussed in \cite{Cas}.} of the Halphen surfaces of index 2 always exist. Because the Halphen surfaces of index 2 are bisection geometries, these surfaces are given by double covers of quartic equations \cite{BM}. By taking the resolvent cubic, the Jacobian fibrations can be constructed \cite{BM}. A Halphen surface of index 2 has a unique double fiber; when the Jacobian fibration is considered, the double fiber becomes a smooth fiber.  
\par This note is structured as follows: After we review the general theory of the Halphen surfaces of index 2 in Section \ref{ssec2.1}, we give explicit constructions of the Halphen surfaces of index 2 with type $I_n$ fibers as described in Section \ref{ssec2.2}. Considering double covers of the Halphen surfaces of index 2 with type $I_n$ fibers that we constructed in section \ref{ssec2.2}, we obtained genus-one fibered K3 surfaces lacking a global section in Section \ref{ssec3.1}. The resulting K3 surfaces have bisection geometries. Furthermore, we show that these K3 genus-one fibrations do not have a global section, but they in fact have a bisection. We show the pullback of a bisection to Halphen surfaces of index 2 gives a bisection to the resulting K3 surfaces. F-theory compactifications on the resulting genus-one fibered K3 surfaces without a section times a K3 surface are discussed in Section \ref{ssec3.2} and Section \ref{ssec3.3}. A discrete $\Z_2$ symmetry and $SU(N)$ gauge symmetry arise in these compactifications, as discussed in Section \ref{ssec3.2}. Additionally, we discuss matter arising from the $A_{n-1}$ singularities in F-theory flux compactifications on these K3 surfaces with $I_n$ fibers times a K3 surface in Section \ref{ssec3.3}. In Section \ref{sec4}, we discuss the Jacobian fibrations of the Halphen surfaces of index 2 and the associated K3 surfaces without a section obtained as double covers of the Halphen surfaces of index 2. We state the concluding remarks in Section \ref{sec5}.

\section{Halphen surfaces of index 2 with $I_n$ fiber}
\label{sec2}
\subsection{Review of general theory of Halphen surfaces of index 2}
\label{ssec2.1}
\par We review the general theory of a class of rational elliptic surfaces, Halphen surfaces of index 2. These surfaces are genus-one fibered, but they do not admit a global section. We introduce this family of rational elliptic surfaces, and show that they indeed lack a global section.
\par Suppose that we have a sextic curve with (at least) nine simple singularities in $\P^2$. (The curve may have more than nine singularities.) We denote the degree 6 polynomial that describes this curve by $f$. The polynomial $f$ can be reducible. We denote nine singularities by $P_i$, $i=1, \cdots, 9$. We mainly consider the situations in which the sextic curve $f=0$ is reducible into lines and conics, or reducible into two cubics with nodes, to construct various examples of Halphen surfaces of index 2. A cubic polynomial in $\P^2$ consists of 10 monomials; therefore, a cubic curve $C$ that passes through nine fixed singular points $P_i$, $i=1, \cdots, 9$, always exists. We denote the equation that describes this cubic curve by $g=0$. We choose this cubic curve to be smooth. 
\par The blow-up of $\P^2$ at nine simple singularities of the sextic curve $f=0$ yields a rational surface. We consider the ratio of the degree 6 polynomial $f$ to the square of the cubic polynomial $g$, $[f: g^2]$; this induces a projection onto $\P^1$. This projection endows the rational surface as the nine-point blow-up of $\P^2$ with a fibration structure. This surface is called the Halphen surface of index 2. The sextic curve $f=0$ defines a divisor that belongs to the complete linear system
\begin{equation}
\label{linear system 1}
|6H-2\, \Sigma_{i=1}^9 P_i|.
\end{equation}
$H$ denotes the line class in $\P^2$. Cubic curve $g=0$ that passes through the nine singularities corresponds to a divisor that belongs to the complete linear system 
\begin{equation}
\label{linear system 2}
|3H-\Sigma_{i=1}^9 P_i|.
\end{equation}
(We have also used $P_i$, $i=1, \cdots, 9$, to denote the exceptional divisors that arise when the nine singularities are blown up in the complete linear systems (\ref{linear system 1}) and (\ref{linear system 2}).)
\par Fiber $F$ of the projection $[f: g^2]$ is given by
\begin{equation}
\label{eq fiber in 2.1}
f+\lambda\, g^2=0,
\end{equation}
where $\lambda$ is a constant. The equation of a fiber (\ref{eq fiber in 2.1}) is a polynomial of degree 6 in $\P^2$. The genus $g(F)$ of a fiber $F$ described by equation (\ref{eq fiber in 2.1}) is given by the genus formula:
\begin{equation}
g(F)=\frac{1}{2}(6-1)(6-2)-\delta=10-9=1.
\end{equation}
where $\delta=9$ is the number of simple singularities of the fiber $f+\lambda g^2=0$. Thus, we conclude that a generic fiber is a smooth genus-one curve when the 9 singularities are blown up, and the projection onto base $\P^1$ given by the map $[f: g^2]$ yields a genus-one fibration. This shows that a Halphen surface of index 2 is genus-one fibered.  
\par We show that the Halphen surfaces of index 2 do not have a section. We denote the fiber class by $F$. Fiber over the point $[a:b]$ is given by
\begin{equation}
b\cdot f-a\cdot g^2=0.
\end{equation}
Particularly, fiber over the origin $[0:1]$ is given by $[f=0]$, and fiber at infinity, $[1:0]$, is given by $[g^2=0]=2\cdot [g=0]$. Thus, we obtain the following linear equivalence:
\begin{equation}
F \sim [f=0] \sim 2\cdot [g=0].
\end{equation}
In particular, it follows that the intersection number of fiber $F$ with any divisor $D$ is a multiple of 2. Because the intersection number of a section with fiber $F$ is 1, this shows that the Halphen surfaces of index 2 do not have a global section. 
\par This argument also shows that the fiber at infinity $[1:0]$ is a double fiber. Exceptional divisors that arise from nine singularities when these singularities are blown up yield bisections. 
\par In Section \ref{ssec2.2}, we explicitly construct sextic curves in $\P^2$, and we construct various Halphen surfaces of index 2 that have type $I_n$ fibers. 

\subsection{Constructions of Halphen surfaces of index 2 with $I_n$ fiber}
\label{ssec2.2}
We choose specific sextic curves $f=0$ in $\P^2$ to construct various Halphen surfaces of index 2 with type $I_n$ fiber. 

\subsubsection{Halphen surface with $I_1$ fiber}
We consider a rational sextic curve $f=0$ with 10 nodes in $\P^2$. We choose any nine nodes, and we consider the blow-up of $\P^2$ at the chosen nine nodes. Consequently, we obtain a Halphen surface of index 2, by considering the projection $[f: g^2]$ onto $\P^1$. As discussed in Section \ref{ssec2.1}, $g=0$ is a cubic that passes through the chosen nine nodes. By construction, the surface has a singular fiber at the origin, which is the rational sextic curve with the one remaining node, i.e., a type $I_1$ fiber. 
\par As discussed in Section \ref{ssec3.1}, when we consider double covers of the resulting Halphen surface with type $I_1$ fiber, we obtain two types of genus-one fibered K3 surfaces without a section: We obtain a K3 surface with two type $I_1$ fibers when the double cover is ramified over a smooth fiber, and we obtain a K3 surface with a type $I_2$ fiber when we consider the double cover of the constructed Halphen surface with type $I_1$ fiber, ramified over the type $I_1$ fiber.
\par Similar arguments as that stated previously apply to other constructions of Halphen surfaces of index 2 with type $I_n$ fibers, $n=2,3,5,6$, that are discussed below. Namely, a double cover of the Halphen surface of index 2 with type $I_n$ fiber that is ramified over a smooth fiber yields a genus-one fibered K3 surface, the singular fibers of which include two type $I_n$ fibers. When we consider a double cover of the Halphen surface of index 2 with type $I_n$ fiber branched over the type $I_n$ fiber at the origin, we obtain a genus-one fibered K3 surface, the singular fibers of which include a type $I_{2n}$ fiber. The resulting K3 surfaces generically lack a global section, and they have bisection geometries, as shown in Section \ref{ssec3.1}.

\subsubsection{Halphen surface with $I_2$ fiber}
\begin{figure}
\begin{center}
\includegraphics[width=\linewidth,height=\textheight,keepaspectratio]{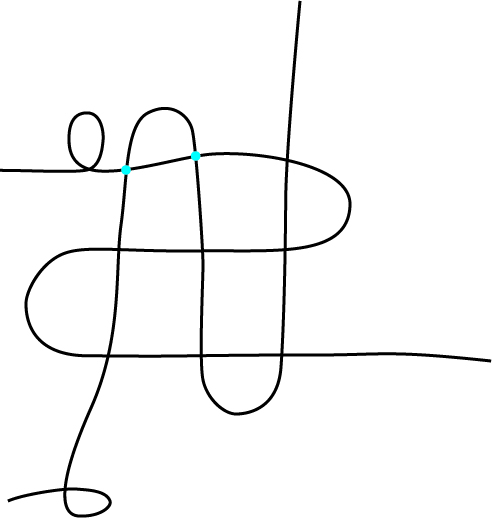}
\caption{\label{fig1}Configuration of two cubics with a node and nine intersection points. The two blue dots represent the two points that are left un-blown up. The remaining nine singularities (two nodes plus seven intersection points) are blown up.}
\end{center}
\end{figure}

We choose two cubics, $f_1, f_2$, in $\P^2$, both of which have one node singularity. The cubic curves $f_1=0$ and $f_2=0$ intersect at nine points. We choose the polynomials $f_1, f_2$ so that these nine intersection points do not include the node singularities that the cubics have. With this choice, the sextic curve $f_1f_2=0$ has 11 singularities in total: nine intersection points of the cubics $f_1=0$ and $f_2=0$ plus the two nodes of the cubics. We leave two points among the nine intersection points of the cubics $f_1=0$ and $f_2=0$, and we consider the blow-up of $\P^2$ at the remaining nine singularity points. See figure \ref{fig1} for two cubics with nodes, and the two points chosen to be left un-blown up.
\par This yields a Halphen surface of index 2. The projection $[f_1f_2 : g^2]$ onto $\P^1$ induces a genus-one fibration. A cubic curve in $\P^2$ with a node is a rational curve (with one singularity) by the genus formula; the genus of a cubic curve in $\P^2$ with a node is
\begin{equation}
\frac{1}{2}(3-1)(3-2)-1=0.
\end{equation}
Therefore, by construction, the resulting Halphen surface has a singular fiber at the origin, which consists of two rational curves meeting at two points (these two points are the points that we chose among the 11 singular points to leave un-blown up); this is a type $I_2$ fiber. 

\subsubsection{Halphen surface with $I_3$ fiber}
We consider three conics $f_1,f_2, f_3$ in $\P^2$. Any two of these conics meet in four points in $\P^2$. Therefore, there are 12 intersection points in total; the sextic curve $f_1 f_2 f_3=0$ has 12 singularities. We leave three points among these 12, and we consider the blow-up of $\P^2$ at the remaining nine intersection points. This yields a Halphen surface. The projection $[f_1f_2f_3 : g^2]$ onto $\P^1$ yields a genus-one fibration.
\par A conic in $\P^2$ is isomorphic to $\P^1$; thus, the resulting Halphen surface has a singular fiber, which consists of three $\P^1$s forming a triangle. (The vertices of this triangle are the three points that we chose among 12 intersection points to leave un-blown up.) This is a type $I_3$ fiber; therefore, the resulting Halphen surface of index 2 has a type $I_3$ fiber at the origin. 

\subsubsection{Halphen surface with $I_5$ fiber}
\begin{figure}
\begin{center}
\includegraphics[width=\linewidth,height=\textheight,keepaspectratio]{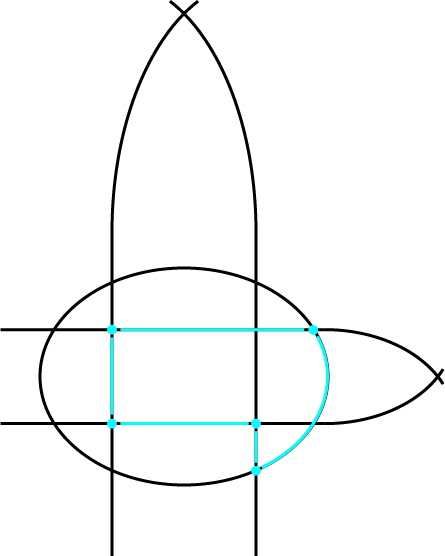}
\caption{\label{fig2}Configuration of four lines and a conic. They have 14 intersection points in total. The five blue dots represent the five points chosen to remain un-blown up. Lines and the conic that pass through these points form a pentagon, as the blue lines show.}
\end{center}
\end{figure}

We consider four lines $f_1, f_2, f_3, f_4$ and a conic $f_5$ in $\P^2$. Two lines intersect at one point and we have $\binom{4}{2}=6$ pairs of lines; the conic and each of the four lines intersect at two points. Therefore, there are 14 intersection points in total. We choose five specific points among these, so that the lines and the conics that pass through the five points form a pentagon. See figure \ref{fig2} for an illustration of this configuration. We consider the blow-up of $\P^2$ at the remaining nine intersection points to obtain a Halphen surface. The projection $[\Pi^5_{i=1}f_i : g^2]$ yields a genus-one fibration. 
\par A line and a conic in $\P^2$ are isomorphic to $\P^1$; therefore, by construction, the resulting Halphen surface of index 2 includes a singular fiber of type $I_5$ at the origin.  

\subsubsection{Halphen surface with $I_6$ fiber}
\begin{figure}
\begin{center}
\includegraphics[width=\linewidth,height=\textheight,keepaspectratio]{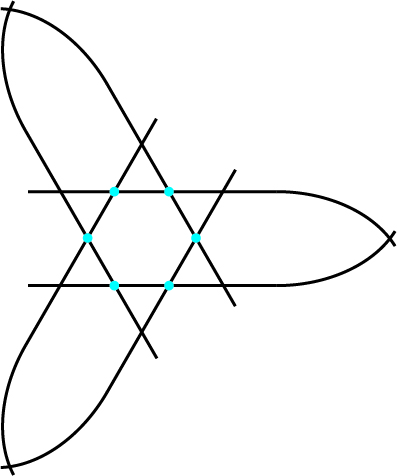}
\caption{\label{fig3}Configuration of six lines and 15 intersections. The six blue dots represent the six points that are left un-blown up. Lines that pass through these six points form a hexagon.}
\end{center}
\end{figure}

We consider six lines $f_i$, $i=1,2,\cdots, 6$, in $\P^2$. We have $\binom{6}{2}=15$ pairs of lines; therefore, there are 15 intersection points in total. We choose the six specific points among these 15 points, so that the lines that pass through the six points form a hexagon. See figure \ref{fig3} for the configuration. The blow-up of $\P^2$ at the remaining nine points yields a Halphen surface. The projection $[\Pi^6_{i=1} f_i : g^2]$ yields a genus-one fibration. By construction, the singular fibers of the resulting Halphen surface include a type $I_6$ fiber at the origin. 

\section{K3 surfaces without a section constructed as double covers of Halphen surfaces of index 2 and F-theory compactifications}
\label{sec3}
\subsection{Constructions of K3 surfaces without a section as double covers of Halphen surfaces}
\label{ssec3.1}
K3 surfaces are obtained by taking double covers of the Halphen surfaces of index 2 that are branched over a fiber. The resulting K3 surfaces are genus-one fibered, but they generically lack a global section to the fibration. When we consider double covers of the Halphen surfaces of index 2, we obtain two different kinds of K3 surfaces, depending on whether a double cover is branched over a smooth fiber or a singular fiber. In either case, the pullback of a bisection of a Halphen surface of index 2 yields a bisection; the resulting K3 surface possesses a bisection to the fibration. In a special situation, bisection of a K3 surface obtained as the pullback of a bisection of a Halphen surface of index 2 splits into two sections. The resulting K3 surface as a double cover of a Halphen surface of index 2 admits a global section for this special case. This case is discussed in Section \ref{sssec3.1.2}. 

\subsubsection{K3 surface without a section as a double cover of Halphen surface ramified over a smooth fiber}
\label{sssec3.1.1}
Fiber of a Halphen surface of index 2 is given by 
\begin{equation}
\label{fiber eq in 3.1.1}
f+a\, g^2=0.
\end{equation}
As stated in Section \ref{sec2}, $f$ is a polynomial of degree 6, so that the sextic curve $f=0$ has nine singular points, and $g=0$ is a smooth cubic that passes through the nine singular points. $a$ in the equation is a constant. For a generic constant $a$, equation (\ref{fiber eq in 3.1.1}) describes a smooth fiber. 
\par We consider a double cover of a Halphen surface of index 2, given by 
\begin{equation}
\label{double cover in 3.1.1}
\tau^2= f+a\, g^2.
\end{equation}
Equation (\ref{double cover in 3.1.1}) is a double cover of a Halphen surface ramified over a sextic curve; thus it yields a K3 surface. The N\'eron--Severi lattice of the resulting K3 surface is generically generated by the smooth fiber, fiber components, and the pullbacks of bisections. The double cover (\ref{double cover in 3.1.1}) is branched over a fiber in the Halphen surface of index 2; because bisection of the Halphen surface of index 2 has the intersection number 2 with fiber, a bisection of the Halphen surface intersects with the fiber as the branching locus at two points. This means that the pullback of a bisection to a K3 surface as the double cover (\ref{double cover in 3.1.1}) is ramified over two points. Thus, the pullback of a bisection of the Halphen surface of index 2 is a double cover of that bisection, which is isomorphic to $\P^1$, ramified over two points. Because the pullback of a bisection is generically irreducible, this argument shows that the pullback of a bisection of the Halphen surface of index 2 is isomorphic to $\P^1$. The pullback of a bisection intersects with the fiber as the branching locus in two points; therefore, it yields a bisection of the K3 surface (\ref{double cover in 3.1.1}). Thus, we conclude that the resulting K3 surface (\ref{double cover in 3.1.1}) generically has only bisections; it does not have a section. 
\par The K3 surface as a double cover (\ref{double cover in 3.1.1}) of a Halphen surface of index 2 has a global section when the pullback of a bisection of the Halphen surface splits into two sections. This occurs when $a=0$ and any one of nine singularities of the polynomial $f$ is a cusp. We discuss this special situation in Section \ref{sssec3.1.2}. 
\par By a reasoning similar to that stated in \cite{KRES}, a K3 surface (\ref{double cover in 3.1.1}) obtained as a double cover of a Halphen surface is equivalent to the quadratic base change of the Halphen surface. This shows that the singular fibers of the resulting K3 surface (\ref{double cover in 3.1.1}) are twice those of the Halphen surface of index 2, when the fiber as the branching locus of the double cover is smooth. See \cite{KRES} for discussion of the configuration of the singular fibers when the quadratic base change is considered. When $a$ takes the value $\infty$, the equation (\ref{double cover in 3.1.1}) becomes 
\begin{equation}
\tau^2=g^2,
\end{equation}
which splits into linear factors as 
\begin{eqnarray}
\tau & = & g \\ \nonumber
\tau & = & -g.
\end{eqnarray}
This corresponds to a situation in which the K3 surface splits into two rational elliptic surfaces. This situation is discussed in \cite{KRES}. We did not consider this situation in this study; we assumed that $a$ is finite. When $a=0$, the fiber $f+a\, g^2=0$ becomes singular. We consider this case in Section \ref{sssec3.1.2} separately. We assume $a\ne 0$ here.
\par The argument stated above shows that the K3 surface (\ref{double cover in 3.1.1}) obtained as a double cover of a Halphen surface of index 2 has twice the number of singular fibers that the Halphen surface has. Particularly, the K3 surfaces obtained as the double covers of the Halphen surfaces constructed as described in Section \ref{ssec2.2} have two type $I_n$ fibers, $n=1,2,3,5,6$.   

\subsubsection{K3 surface without a section as a double cover of Halphen surface ramified over a singular fiber}
\label{sssec3.1.2}
We consider a double cover of a Halphen surface of index 2 that is ramified over a singular fiber. This corresponds to the case where the constant $a$ takes the value 0 in equation (\ref{double cover in 3.1.1}). For this case, the double cover of a Halphen surface of index 2 is given by
\begin{equation}
\label{eq Hal 3.1.2}
\tau^2= f.
\end{equation}
This is the quadratic base change of the Halphen surface, where a ramification occurs over the singular fiber $f=0$. Similar reasoning as that given in Section \ref{sssec3.1.1} shows that the resulting K3 surface (\ref{eq Hal 3.1.2}) has a bisection, but it generically does not have a section. When any one of nine singularities that the polynomial $f$ possesses is a cusp, a bisection of the Halphen surface, which arises as an exceptional curve when the cusp is blown up, is tangent to the branch; therefore, the pullback of that bisection splits into two sections \footnote{When $a$ is nonzero, the polynomial $f+a\, g^2$ generally does not have a cusp.}. For this special situation, the resulting K3 surface has a section. We did not consider such a situation in this study.
\par We constructed the Halphen surfaces of index 2 with type $I_n$ fibers, $n=1,2,3,5,6$, in Section \ref{ssec2.2}. When we consider the double covers (\ref{eq Hal 3.1.2}) of these Halphen surfaces, the double covers are ramified over the type $I_n$ fiber. This means that the corresponding quadratic base change is ramified over type $I_n$ fiber, and two type $I_n$ fibers collide and they are enhanced to a type $I_{2n}$ fiber \cite{SchShio}  \footnote{The situation in which the quadratic base change is ramified over a singular fiber, and two identical singular fibers collide and they are enhanced to another type of singular fiber, is discussed in the context of F-theory compactification in \cite{KRES}.}. This shows that the resulting K3 surfaces (\ref{eq Hal 3.1.2}) as the double covers of the Halphen surfaces with type $I_n$ fibers, $n=1,2,3,5,6$, have type $I_{2n}$ fibers.  

\subsection{Gauge symmetries in F-theory compactifications}
\label{ssec3.2}
It was shown in Section \ref{ssec3.1} that genus-one fibered K3 surfaces obtained as double covers of Halphen surfaces of index 2 lack a global section, but they have a bisection. Thus, a discrete $\Z_2$ symmetry arises in F-theory compactifications on spaces built as the direct products of these K3 surfaces times a K3 surface. 
\par The singular fibers of genus-one fibered K3 surfaces without a section obtained in Section \ref{sssec3.1.1} include two type $I_n$ fibers, $n=2,3,5,6$, therefore, $SU(n)^2$ gauge groups arise in F-theory compactifications. Thus, we conclude that gauge groups that arise in F-theory compactifications on genus-one fibered K3 surfaces without a section obtained in Section \ref{sssec3.1.1} times a K3 contain the following factor:
\begin{equation}
SU(n)^2 \times \Z_2.
\end{equation}
\par The singular fibers of genus-one fibered K3 surfaces without a section obtained in Section \ref{sssec3.1.2} include type $I_m$ fibers, $m=2,4,6,10,12$. Gauge groups that arise in F-theory compactifications on genus-one fibered K3 surfaces without a section obtained in Section \ref{sssec3.1.2} times a K3 contain the following factor:
\begin{equation}
SU(m) \times \Z_2.
\end{equation}

\subsection{Matter fields in F-theory compactifications on the resulting K3 surfaces}
\label{ssec3.3}
F-theory compactification on a space constructed as the direct product of K3 surfaces yields a four-dimensional theory with $N=2$ supersymmetry. By including four-form flux \cite{BB, SVW, W, GVW, DRS}, the supersymmetry of the theory reduces to $N=1$. Hypermultiplets split into vector-like pairs when the four-form flux is turned on. We do not discuss the cancellation of the tadpole with flux \cite{SVW} in this note. We can only state that vector-like pairs could arise, owing to the tadpole. Matter spectra in F-theory flux compactifications on the direct products of K3 surfaces, where one of the two K3 surfaces in the product lacks a global section, are discussed, for example, in \cite{K, K2}. 
\par We discuss the vector-like pairs that can arise from the $A_{n-1}$ singularities corresponding to type $I_n$ fibers in F-theory compactifications with flux on K3 surfaces without a section, which we obtained as described in Section \ref{ssec3.1} as double covers of the Halphen surfaces, times a K3 surface. Under the rank 1 enhancement
\begin{equation}
A_{n-2}\subset A_{n-1},
\end{equation}
the adjoint representation ${\bf n^2-1}$ of $SU(n)$ decomposes into irreducible representations of SU($n-1$) as \cite{Sla}:
\begin{equation}
{\bf n^2-1}=[{\bf (n-1)^2-1}]+{\bf n-1}(\, \ytableausetup{boxsize=.6em}\ytableausetup
{aligntableaux=center}\begin{ytableau}
 \
\end{ytableau} \,)+\overline{\bf n-1}+{\bf 1}.
\end{equation}
Thus, the adjoints ${\bf (n-1)^2-1}$ of $SU(n-1)$ arise on the 7-branes without fluxes. The vector-like pairs, ${\bf n-1}\, +\, \overline{\bf n-1}$, can arise from the $A_{n-1}$ singularities with flux. ($n=3,4,5,6,10,12$)

\section{Jacobian fibrations of Halphen surfaces and associated K3 surfaces}
\label{sec4}
\par The Halphen surfaces of index 2 have Jacobian fibrations. The locations and the types of the singular fibers of the Jacobian fibration are identical to those of Halphen surface, except the multiple fiber. The Jacobian fibration has a global section, therefore it does not have a multiple fiber. The double fiber at infinity of a Halphen surface of index 2 becomes a smooth fiber when we take the Jacobian fibration. 
\par Halphen surfaces of index 2, and generic genus-one fibered K3 surfaces obtained as double covers of the Halphen surfaces of index 2 are bisection geometries. Therefore, these surfaces can be described as double covers of the quartic polynomials \cite{BM}. The Jacobian fibrations of these surfaces are obtained by taking a resolvent cubic of the quartic polynomials \cite{BM}.

\section{Conclusions}
\label{sec5}
In this note, we used a special class of rational elliptic surfaces lacking a global section, Halphen surfaces of index 2, to construct genus-one fibered K3 surfaces without a global section. The resulting K3 surfaces are bisection geometries. Therefore, a discrete $\Z_2$ gauge symmetry arises in F-theory compactifications on these genus-one fibered K3 surfaces times a K3 surface.
\par K3 surfaces without a section are obtained when we consider the double covers of the Halphen surfaces of index 2. We showed in this study that the resulting K3 surfaces have bisections, but they do not have a section. We obtain two types of K3 surfaces, depending on whether the ramification of the double covers occurs along a smooth fiber or a singular fiber. When the double cover is ramified over a smooth fiber, the resulting K3 surface has twice the singular fibers that the Halphen surface has. When we consider the double cover of this type, we obtain K3 surfaces without a section with type $I_n$ fibers. When the ramification occurs along a singular fiber, two fibers of type $I_n$ fibers collide, and they are enhanced to a type $I_{2n}$ fiber. The resulting K3 surfaces have type $I_{2n}$ fibers in this situation. 
\par $SU(n)^2\times \Z_2$, $n=2,3,5,6$ gauge groups arise in F-theory compactifications on the K3 surfaces as the double covers of the Halphen surfaces of the first type times a K3 surface; $SU(m)\times \Z_2$, $m=2,4,6,10,12$, gauge symmetries arise in F-theory compactification on the K3 surface as double covers of the Halphen surfaces of the second type times a K3 surface.

\section*{Acknowledgments}

We would like to thank Shun'ya Mizoguchi and Shigeru Mukai for discussions. This work is partially supported by Grant-in-Aid for Scientific Research {\#}16K05337 from The Ministry of Education, Culture, Sports, Science and Technology of Japan.

\end{document}